\documentclass[10pt,draft]{dis03}
\usepackage{epsf,amsmath}

\begin{document}

\title{PHENOMENOLOGY OF A SU(2) TRIPLET HIGGS
\thanks{This work has been supported by the UK Particle Physics and Astronomy Research Council (PPARC) (Posdoctoral Fellowship: PPA/P/S/1999/00446).}}

\author{AGUST{\'\i}N SABIO VERA\\\\
Cavendish Laboratory,
  University of Cambridge, Madingley Road,\\ CB3 0HE, Cambridge, U.K.
\vspace{0.3cm}}

\maketitle

\begin{abstract}
\noindent 
We study the Renormalization Group (RG) evolution 
of the couplings in a model with a real SU(2) triplet in the Higgs sector. 
Insisting that the model remain valid up to 1~TeV we show that 
it is possible for there to be no light Higgs bosons without
any otherwise dramatic deviation from the physics of the Standard Model.
\end{abstract}

\section{Introduction}
In this contribution we extend the study in Ref. \cite{Forshaw:2001xq} of an 
extension of the Standard Model in which a real scalar $SU(2)$ triplet with 
zero hypercharge is added to the usual scalar $SU(2)$ doublet. For details 
see Ref. \cite{Forshaw:2003kh}. The scalar potential of the model in terms of the 
usual Standard Model Higgs, $\Phi_1$, and the new triplet, $\Phi_2$, reads
\begin{eqnarray}
V_0(\Phi_1 , \Phi_2) &=&\mu_1^2 ~|\Phi_1|^2 + \frac{\mu_2^2}{2} 
~|\Phi_2|^2 + \lambda_1 ~|\Phi_1|^4+ \frac{\lambda_2}{4} ~|\Phi_2|^4 
+ \frac{\lambda_3}{2} ~|\Phi_1|^2 ~|\Phi_2|^2 \nonumber\\
&+& \lambda_4 \,  {\Phi_1}^\dagger \sigma^\alpha \Phi_1 ~ {\Phi_2}_\alpha,\nonumber
\end{eqnarray}
where $\sigma^\alpha$ are the Pauli matrices.
The expansion of the field components is
\begin{eqnarray}
\Phi_1 = \left(\begin{array}{c}\phi^+ \\
\frac{1}{\sqrt{2}}\left(h_c^0 + h^0 + i \phi^0\right)\end{array}\right)_
{Y=1},
~~\Phi_2 = \left(\begin{array}{c}\eta_1 \\
\eta_2 \\ \eta_c^0 + \eta^0    \end{array}\right)_{Y=0}\nonumber,
\end{eqnarray}
with $\eta^\pm = ( \eta_1 \mp i \eta_2) / \sqrt{2}$ and $\phi^0$ is the
Goldstone boson which is eaten by the $Z^0$. The model violates custodial 
symmetry at tree level giving a prediction of 
$\rho ~=~ 1 + 4 \left(\frac{\eta^0_c}{h^0_c}\right)^2$
for the $\rho$-parameter. In the neutral Higgs sector we have two CP-even states which mix 
with angle $\gamma$. The mass eigenstates $\{H^0, N^0\}$ are defined by 
\begin{eqnarray}
\left(\begin{array}{c}H^0\\N^0\end{array}\right) &=& 
\left(\begin{array}{cc}\cos{\gamma}&-\sin{\gamma}\\
\sin{\gamma}&\cos{\gamma}\end{array}\right)
\left(\begin{array}{c}h^0\\\eta^0\end{array}\right).\nonumber
\end{eqnarray}
In the charged Higgs sector the mass eigenstates $\{g^\pm, h^\pm\}$ are 
\begin{eqnarray}
\left(\begin{array}{c}g^\pm\\h^\pm\end{array}\right) &=& 
\left(\begin{array}{cc}\cos{\beta}&-\sin{\beta}\\
\sin{\beta}&\cos{\beta}\end{array}\right)
\left(\begin{array}{c}\phi^\pm\\\eta^\pm\end{array}\right).\nonumber
\end{eqnarray}
The $g^\pm$ are the Goldstone bosons corresponding to $W^\pm$ and,
at tree level, the mixing angle is $\tan{\beta} = 2 \frac{\eta^0_c}{h^0_c}$.
The precision electroweak data constrain $\beta$ to be smaller
than about $4^\circ$ \cite{Forshaw:2001xq}.

\section{The beta-functions}

In Ref. \cite{Forshaw:2003kh} the beta functions for the couplings were 
calculated using the one-loop effective potential 
\cite{effective} 
with $\overline{\rm MS}$ renormalization in 't~Hooft-Landau gauge and 
the anomalous dimensions for $h^0$ and $\eta^0$, the results read
\begin{eqnarray}
\beta_{\mu_1} &=&
\frac{1}{16 \pi^2}\,\left(6 \, \lambda_4^2 + 12 \, \lambda_1 \mu_1^2 
+ 3 \, \lambda_3 \, \mu_2^2\right)
+ \frac{1}{8 \pi^2}\,\left(3\,h_t^2 
-\frac{9}{4}\,g^2 - \frac{3}{4}\,{g'}^2\right)\,\mu_1^2,\nonumber\\
\beta_{\mu_2} &=& 
\frac{1}{16 \pi^2}\,\left(4\,\lambda_4^2 + 4\, \lambda_3 \, \mu_1^2 + 10 \,
\lambda_2 \, \mu_2^2 \right)
-\frac{3}{4 \pi^2}\,g^2 \, \mu_2^2 ,\nonumber\\
\beta_{\lambda_1} &=& \frac{1}{8 \pi^2}\left(\frac{9}{16}\,g^4 - 3\,{{h_t}}^4 
+12 \, \lambda_1^2 + \frac{3}{4}\,\lambda_3^2 + \frac{3}{8}\,g^2\,{g'}^2
+ \frac{3}{16}\,{g'}^4\right) \nonumber\\
&+&\frac{1}{4 \pi^2}\,\left(3\,h_t^2 
-\frac{9}{4}\,g^2 - \frac{3}{4}\,{g'}^2\right)\,\lambda_1,\nonumber\\
\beta_{\lambda_2} &=& \frac{1}{8\pi^2}
\left(6\,g^4 + 11\,\lambda_2^2 + \lambda_3^2 \right)
-\frac{3}{2 \pi^2}\,g^2 \, \lambda_2,\nonumber\\
\beta_{\lambda_3} &=& \frac{1}{8\pi^2}
\left(3 \, g^4 + 6\,\lambda_1 \, \lambda_3 + 5 \, \lambda_2 \, \lambda_3 
+ 2 \, \lambda_3^2\right)
+\frac{3 \lambda_3}{8 \pi^2}\,\left(h_t^2 
-\frac{11}{4}\,g^2 - \frac{1}{4}\,{g'}^2\right),\nonumber\\
\beta_{\lambda_4}&=& \frac{1}{4\pi^2}\,\lambda_4 \, 
\left(\lambda_1 + \lambda_3 \right)+\frac{3}{32 \pi^2}\,\left(4\,h_t^2 
- 7\,g^2 - \,{g'}^2\right)\,\lambda_4.\nonumber
\end{eqnarray}
In the gauge and top quark 
sector the beta functions for the $U(1)$, $SU(3)$ and Yukawa 
couplings are the same as in the Standard Model and only the $SU(2)$ 
coupling is modified due to the extra Higgs triplet in the adjoint 
representation, i.e. $\beta_g = - \frac{5}{32 \pi^2}\,g^3$. 

Working with the tree-level effective potential with couplings
evolved using the one-loop $\beta$ and $\gamma$ functions we are able to 
resum the leading logarithms to all orders in the effective potential. 
To carry out the RG analysis we first introduce the 
parameter $t$, related to $\mu$ through  $\mu (t) = m_Z \exp{(t)}$. We 
perform evolution starting at $t=0$.
The RG equations are coupled differential equations in the 
set $\left\{g_s, ~g, ~g',~h_t,~\mu_1,~\mu_2,~\lambda_1,~\lambda_2,~\lambda_3
,~\lambda_4\right\}$. 
We choose rather to use the set 
$\left\{\alpha_s,~m_Z,~\sin^2{\theta_W},~m_t,~m_{h^\pm},~m_{H^0},~m_{N^0},~v,~
\tan{\beta},~\tan{\gamma} \right\}$.

Within the accuracy to which we are working, the values of the 
couplings at $t=0$ can be obtained from the input set
using the appropriate tree-level expressions. Making use of the vacuum 
conditions and the notation $h^0_c \equiv v$ and 
$\eta^0_c ~\equiv~ \frac{v}{2}\,\tan{\beta}$ we can write the 
tree level masses as  
\begin{eqnarray}
m_Z^2 &=&\frac{1}{4}\,v^2\,\left( g^2 + {g'}^2 \right), ~~m_W^2 ~=~ \frac{1}{4}\,g^2\,{v}^2\,\left(1+\tan^2{\beta}\right), \nonumber\\
m_t^2 &=& \frac{1}{2}{h_t}^2\,{v}^2, ~~m_{h^\pm}^2 ~=~ v\,\lambda_4 \,\left(\cot{\beta}+\tan{\beta}\right),
\nonumber\\ 
m_{H^0}^2 &=& v\,\left\{2\,v\,\lambda_1 +
\left(\lambda_4 - \frac{1}{2}\,v \,\lambda_3\, \tan{\beta}\right)
\,\tan{\gamma}\right\}, \nonumber\\
m_{N^0}^2 &=& v\, \lambda_4\,\left(\cot{\beta}-\tan{\gamma}\right)
+\frac{1}{2}\,v^2\,\tan{\beta}\left(\lambda_2\,\tan{\beta}+
\lambda_3\, \tan{\gamma}\right),\nonumber
\end{eqnarray}
with $\tan{(2\,\gamma)} = \frac{2\,\tan{\beta}\,\left(-2\,\lambda_4 + 
v\,\lambda_3\,\tan{\beta}\right)}{2\,\lambda_4 - 4 \, v \, \lambda_1\,
\tan{\beta}+v\,\lambda_2\,\tan^3{\beta}}$. 
Inverting these relations we can fix the $t=0$ boundary
conditions for the subsequent evolution. To ensure that the system remains in 
a local minimum we impose the condition 
that the squared masses should remain positive.
We also impose that the couplings remain
perturbative, insisting that 
$|\lambda_i (t)| < 4 \pi$ for $i={1,2,3}$ and $|\lambda_4| < 4 \pi v$.
We run the evolution from $t = 0$ to $t_{\rm max}= \log{(\Lambda/m_Z)}$, with 
$\Lambda = 1$ TeV.

\section{Results in the non-decoupling regime}

In the non-decoupling regime the triplet cannot be arbitrarily heavy. 
For this case in Fig.~1 we show the range 
of Higgs masses allowed when there is no mixing in the neutral 
Higgs sector, $\gamma = 0$, for $\beta = 0.04$. Such a value is 
interesting because it allows a rather heavy lightest Higgs 
(e.g. for $\beta = 0.04,\; m_{H^0}>150$~GeV and for 
$\beta = 0.05, \; m_{H^0}>300$~GeV)~\cite{Forshaw:2001xq}.
The strong correlation between the $h^\pm$ and $N^0$ masses arises in order
that $\lambda_2$ remain perturbative. The upper bound on the triplet Higgs masses 
($\approx 550$~GeV) comes
about from the perturbativity of $\lambda_3$ whilst that on $H^0$ 
($\approx 520$~GeV) comes
from the perturbativity of $\lambda_1$.  
The hole at low masses is due to vacuum stability. 
\begin{figure}[!thb]
\label{B004G0}
\vspace*{5.0cm}
\begin{center}
\includegraphics{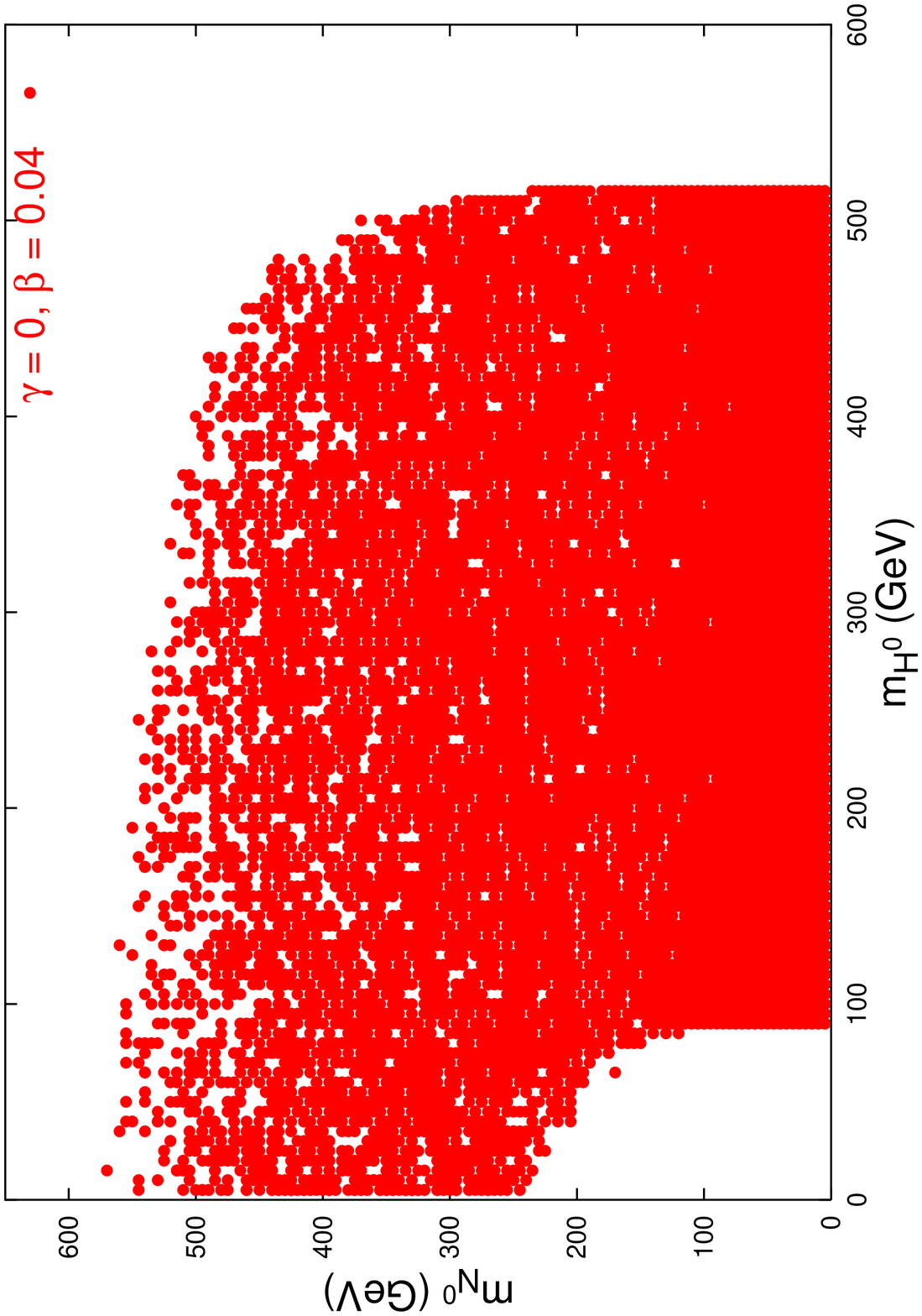}
\includegraphics{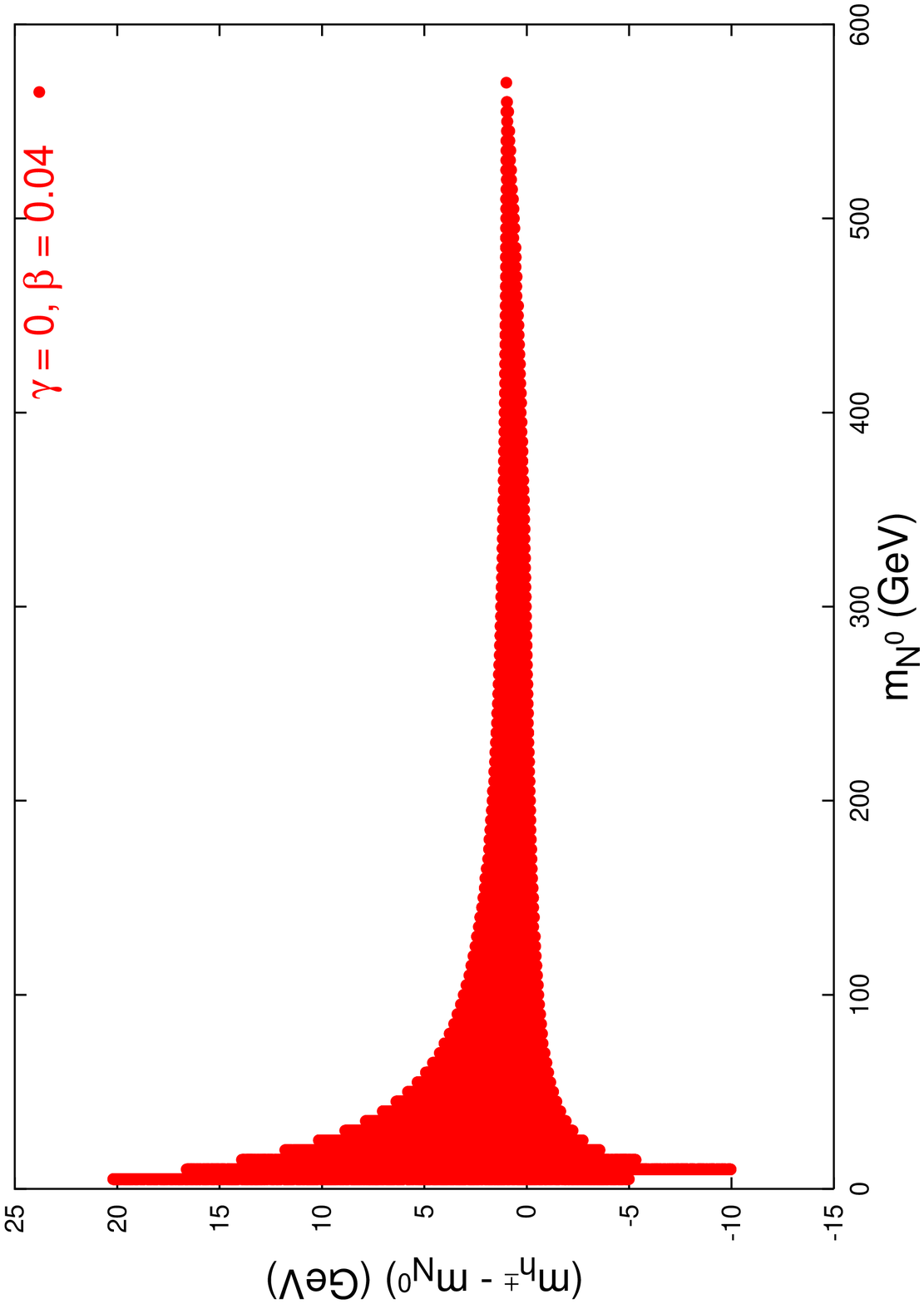}
\caption[*]{Allowed values of scalar masses for $\gamma=0$}
\end{center}
\end{figure}

The $\gamma = 0.1$ case has been considered in Ref. \cite{Forshaw:2003kh} and 
the bounds are similar to the no-mixing case. 
In the maximal mixing case of $\gamma = \pi/4$ the bounds are more 
democratic. The largeness of $\tan(2 \gamma)$ can be arranged either by tuning $2 v\lambda_1
\approx \lambda_4/\beta$ or by having small enough $\lambda_1$ and 
$\lambda_4$. In the former case, all masses are approximately degenerate. 
In the latter case, which corresponds to
light masses, the degeneracy is lifted.
The bounds for $\gamma > \pi/4$ are very similar to those for 
$(\pi/2 - \gamma)$ on interchanging $N^0$ and $H^0$. 
For $\beta < 0.04$ and small $\gamma$ but away from
the decoupling regime the allowed regions are very similar to those 
for $\beta=0.04$. 
For larger $\gamma$, the mass bounds are again as for larger $\beta$. 

\section{The decoupling limit}
For $\beta = \gamma = 0$ there is no doublet-triplet mixing and no bound on 
the triplet
mass. This is a special case of the more general decoupling scenario,
which occurs when $|\beta + \gamma| \ll \beta$, where the triplet decouples 
from the doublet. For small mixing angles, the triplet Higgs has mass squared
$\sim \lambda_4 v / \beta$ and it is possible to have 
$\lambda_4 \sim v$ by keeping $\mu_2^2$ large. In this case 
$\beta+\gamma \approx 0$ \cite{Forshaw:2003kh}. This is the decoupling
limit in which the triplet mass lies far above the mass of the
doublet and the low energy model looks identical to the Standard Model. 

\section{Conclusions}
We have computed the one-loop beta functions for the scalar couplings
in an extension to the Standard Model which contains an additional
real triplet Higgs. Through considerations of perturbativity of the
couplings and vacuum stability we have identified the
allowed masses of the Higgs bosons in the non-decoupling regime \cite{Forshaw:2003kh}. In the
decoupling regime, the model tends to resemble the Standard Model.
 The near degeneracy of the triplet Higgs
masses ensures that,
at least for small $\gamma$, the quantum corrections to the $T$ parameter are 
negligible (the $S$ parameter vanishing since the triplet has zero 
hypercharge) \cite{Forshaw:2001xq}. This means that the lightest Higgs
boson can be heavy as a result of the compensation arising
from the explicit tree-level violation of custodial symmetry and it is 
possible to be in a regime where all the Higgs bosons are heavy without
any dramatic deviation from the physics of the Standard Model.

\section*{Acknowledgements}
I would like to thank Jeff R. Forshaw and Ben E. White, my collaborators in 
this work, and the participants of DIS 2003, in particular, Wilfried Buchmueller and Stefan 
Schlenstedt for their interest in our results.

\end{document}